\documentstyle[aps]{revtex}
\def \be   {\begin{equation}}
\def \ee   {\end{equation}}
\def \l {\label}
\begin{document}
\input epsf
\baselineskip=25pt
\title{Discrete scalar fields and general relativity}
\author{Manoelito M de Souza}
\address{Universidade Federal do Esp\'{\i}rito Santo - Departamento de
F\'{\i}sica\\29065.900 -Vit\'oria-ES-Brasil}
\date{\today}
\maketitle
\begin{abstract}
\noindent The physical meaning, the properties  and the consequences of a
discrete scalar field are discussed; limits for the validity of a continuous
mathematical description of fundamental physics are a natural outcome of
discrete fields with discrete interactions. The discrete scalar field is
ultimately the gravitational field of general relativity, necessarily, and
there is no place for any other fundamental scalar field, in this context.
\end{abstract}
\begin{center}
PACS numbers: $04.20.Cv\;\; \;\; 04.30.+x\;\; \;\;04.60.+n$
\end{center}
\section{Introduction}
Although it is considered that a scalar field has not been observed in nature
as a fundamental field its use as such is very frequent in the modern
literature, particularly in elementary particles, field theory and cosmology.
Here we will apply to the scalar field the concepts and results developed in
the reference \cite{paperI}, referred here as the paper I, where the concept of
a discrete field was introduced and its wave equation and its Green's function
discussed. The standard field and its formalism, which for a distinction, we
always append the qualification continuous, are retrieved from an integration
over the discrete-field parameters. Remarkable in the discrete field is that it
has none of the problems that plague the continuous one so that the meaning and
origin of these problems can be left exposed on the passage from the discrete
to the continuous formalism. Although the motivations for the introduction of a
generic discrete field in paper I have being made on pure physical grounds of
causality, a deeper discussion about its physical interpretation have been left
for subsequent papers on specific fields. This discussion will be retaken here
with the simplest structure of a field, the scalar one. It would be a too easy
posture to see the discrete scalar field as just an ancillary mathematical
construct devoid of any physical meaning, and this vision could be re-enforced
with the discrete  field being just a point propagating with the speed of
light. The idea of a point-like field may sound weird at a first sight but this
represents the same symmetry of quantum field theory where fields and sources
are equally treated as quantized fields. Here it is seen from a reversal and
classical perspective. Besides, point-like objects was never a novelty in
physics and one of the major motivations of the, nowadays so popular, string
theory is of avoiding \cite{Polchinski} infinities and acausalities in the
fields produced by point sources, a problem that do not exist for the discrete
field, according to the reference \cite{hep-th/9610028}. On the other hand, it
could look peculiar, considering the still doubtful existence of a fundamental
scalar field, that it be the one picked for discussing attribution of physical
reality to a discrete field. This usual strategy of considering the simplest
problems first, even if just as a  preliminary academical exercise, produced
however, unexpected bonus. It turns out from the discrete field analysis that
the scalar field is, necessarily, the gravitational field of general
relativity, whose character of a second-rank tensor is assured by the way the
scalar field is attached to the definition of the metric tensor. After decoding
the physical meaning of the scalar field sources one is led to the unavoidable
conclusion that there is no place, in this context, for the existence of any
other fundamental scalar field. This has deeper implications, discussed at the
end of the paper.

This paper is structured in the following way. Section II, on the sake of a
brief review of the mathematical definition of discrete fields, is a recipe on
how to pass from a continuous to a discrete formalism, and vice-versa. The
discrete scalar field, its wave equation, its Lagrangian and its energy tensor
are discussed in Section III and the consequences of discrete interactions for
the mathematical description of the physical world in Section IV. Calculus
(integration and differentiation) which is based on the opposite idea of
smoothness and continuity, has its full validity for describing dynamics
restricted then to a very efficient approximation  in the case of a high
density of interaction points, as turns out to be the great majority of cases
of physical interest. This seems to be an answer to the Wigner's pondering
\cite{Wigner} about the reasons behind the unexpected effectiveness of
mathematics on the physical description of the world. It is argued in Section
V, after analysing the results of the Section IV, that the scalar field must
necessarily describe the gravitational interaction of general relativity.
Section VI brings the conclusions.

\section{From continuous to discrete}

For a concise introduction of the discrete-field concept it is convenient to
replace the Minkowski spacetime flat geometry by a conical projective one of an
embedding (3+2) flat spacetime:
\be
\l{def}
\{x\in R^4\}\Rightarrow\{x,x^5\in R^5{\big|}(x^5)^2+x^2=0\},
\ee
where $x\equiv({\vec x},t)$ and $x^2\equiv\eta_{\mu\nu}x^{\mu}x^{\nu}=|{\vec
x}|^2-t^2.$ So a change $\Delta x^5$ on the fifth coordinate, allowed by the
constraint $(\Delta x^5)^2+(\Delta x)^2=0,$   is a Lorentz scalar that can be
interpreted as a change $\Delta\tau$ on the proper-time  of a physical object
propagating across an interval $\Delta x:\quad \Delta x^5=\Delta\tau.$

The constraint
\be
\l{hcone}
(\tau-\tau_{0})^2+(x-x_{0})^2=0
\ee
defines a double hypercone with vertex at $(x_{0},\tau_{0}),$ whilst
\be
\l{hplane}
(\tau-\tau_{0})+f_{\mu}(x-x_{0})^{\mu}=0
\ee
defines a family of hyperplanes tangent to the double hypercone and labelled by
their normal\footnote{The eq. (\ref{hplane}) can be written in $R^5$ as
$f_{M}\Delta x^{M}=0,\;\;M=1,2,3,4,5$ with $f_{M}=(f_{\mu},1)$} $f_{\mu}$, a
constant four-vector. The intersection of the double hypercone with a
hyperplane defines a $f$-generator tangent to $f^{\mu}$
($f^{\mu}:=\eta^{\mu\nu}f_{\nu}).$ A discrete field is a field defined with
support on this intersection (extended causality) in contraposition
\cite{paperI} to the continuous field, defined with support on a hypercone
(local causality):
\be
\l{df}
\phi_{f}(x,\tau):=\phi(x,\tau){\Big|}_{{\Delta\tau+f.\Delta
x=0}\atop{\Delta\tau^2+\Delta x^2=0}}:=\phi{\Big|}_{f}.
\ee
The symbol ${\big|}_{f}$ is a short notation for the double constraint in the
middle term of eq. (\ref{df}). The constraint (\ref{df}) induces the
directional derivative (along the fibre $f$, the hypercone $f$-generator)
\be
\l{dd}
\nabla_{\mu}\phi_{f}(x,\tau):=(\partial_{\mu}-f_{\mu}\partial_{\tau})\phi_{f}(x,\tau).
\ee

An action for a discrete scalar field is
\be
\l{da1}
S_{f}=\int
d^5x{\Big\{}\frac{1}{2}\eta^{\mu\nu}\nabla_{\mu}\phi_{f}(x,\tau)\nabla_{\nu}\phi_{f}(x,\tau)-\chi\phi_{f}(x,\tau)\rho(x,\tau){\Big\}},
\ee
where $d^5x=d^4xd\tau$, $\chi$ is a coupling constant and $\rho(x,\tau)$ is the
source for the scalar field. There can be no mass term in a discrete-field
Lagrangian because it would imply on a hidden breaking of the Lorentz symmetry
with non-propagating discrete solutions of the  field equations. In other words
no physical object could be described by such a Lagrangian with an explicit
mass term. Nevertheless, as discussed  in paper I, the action (\ref{da1}) still
describes both, massive and massless fields. The mass of a massive discrete
field is implicit on its propagation with a non-constant proper-time. Eq.
(\ref{da1}) is a scale-free action with its element volume $d^5x$ hiding the
actual (1+1)-Lagrangian of a discrete field, massive or not, on a fibre $f$; a
mass term would break its conformal symmetry \cite{paperI}.

Then the field equation and the tensor energy for a discrete field are,
respectively,
\be
\l{dfe}
\eta^{\mu\nu}\nabla_{\mu}\nabla_{\nu}\phi_{f}(x,\tau)=\chi\rho(x,\tau),
\ee
\be
\l{det}
T^{\mu\nu}_{f}=\nabla^{\mu}\phi_{f}\nabla^{\nu}\phi_{f}-\frac{1}{2}\eta^{\mu\nu}\nabla^{\alpha}\phi_{f}\nabla_{\alpha}\phi_{f}.
\ee
They must be compared to the standard expressions for the continuous field:
\be
\l{usfe}
(\eta^{\mu\nu}\partial_{\mu}\partial_{\nu}-m^2)\phi(x)=\chi\rho(x),
\ee
\be
\l{ts}
T^{\mu\nu}(x)=\partial^{\mu}\phi\partial^{\nu}\phi-\frac{1}{2}\eta^{\mu\nu}\partial^{\alpha}\phi\partial_{\alpha}\phi
\ee
which can be obtained from the action
\be
\l{a}
S=\int
d^{4}x{\Big\{}\frac{1}{2}\eta^{\mu\nu}\partial_{\mu}\phi_{f}\partial_{\nu}\phi_{f}-\frac{m^2}{2}\phi^2-\chi\phi(x)\rho(x){\Big\}},
\ee
 So, the passage from a continuous to a discrete field formalism can be
summarized in the following schematic recipe (the arrows indicate
replacements):
\be
\l{recipe}
\cases{\{x\}\Rightarrow\{x,x^5\};\cr
\phi(x)\Longrightarrow \phi(x,\tau){\Big |}_{f};\cr
\partial_{\mu}\Rightarrow\nabla_{\mu},\cr}
\ee
accompanied by a dropping of the mass term from the Lagrangian. Moreover a
discrete field requires discrete sources \cite{paperI}. The continuous
$\rho(x)$ is replaced by a discrete set of point-like sources $\rho(x,\tau)$.
Any apparent continuity is reduced to a question of scale in the observation.
$\rho(x,\tau)$ is, like  $\phi_{f}(x,\tau)$, a discrete field defined on a
hypercone generator too, which just for simplicity, is  not being considered
here. This is a symmetry between fields and sources: they are all discrete
fields, and the current density of one is the source of the other.

In the opposite-direction passage, from discrete to continuous, the continuous
field and its field equations are recuperated in terms of effective average
fields smeared over the hypercone
\be
\label{s}
\Phi(x,\tau)=\frac{1}{2\pi}\int d^{4}f\delta(f^2)\Phi_{f}(x,\tau).
\ee

\section{The discrete scalar field}

Comparing the actions of eqs. (\ref{da1}) and (\ref{a}) one should observe that
the first one contains explicit manifestations only of the constraint
(\ref{hplane}) through the use of the directional derivatives (\ref{dd}), but
not of the constraint (\ref{hcone}). This one is only dynamically introduced
through the solutions of the field equation, like it happens also (local
causality) in the standard formalism of continuous fields \cite{Jackson}. As a
matter of fact all the information contained in the new action (\ref{da1}) can
be incorporated in the old action (\ref{a}), without the mass term, with the
simple inclusion of the constraint (\ref{hplane})
\be
\l{ar}
S_{P}=\int d^{4}xd\tau\delta(\Delta\tau+f.\Delta
x){\Big\{}\frac{1}{2}\eta^{\mu\nu}\partial_{\mu}\phi\partial_{\nu}\phi-\chi\phi(x,\tau)\rho(x,\tau){\Big\}},
\ee
as the very  restriction to the hyperplane (\ref{hplane}) by itself implies on
the whole recipe (\ref{recipe}). $P$ in (\ref{ar}) stands for any generic fixed
point, the local hypercone vertex: $P=(x_{0},\tau_{0}),$
$\Delta\tau=\tau-\tau_{0}$ and $\Delta x=x-x_{0}.$ Local causality, dynamically
implemented through the field equations, imply that the field propagates on a
hypercone (the lightcone, if a massless field) with vertex on $P,$  which is an
event on the world  line of $\rho(x,\tau)$.

Whereas there is no restriction on $\rho(x)$ for a continuous field, for a
discrete one, as already mentioned, it must be\cite{paperI} a discrete set of
point sources. A continuously extended source would not be consistent as it
would produce a continuous field. The source of a discrete scalar field is
given by
\be
\l{rof}
\rho(x,t_{x}=t_{z})=q(\tau_{z})\delta^{(3)}({\vec x}-{\vec
z}(\tau_{z}))\delta(\tau_{x}-\tau_{z}),
\ee
where $z(\tau)$ is its world  line parameterized by its proper time $\tau$;
$q(\tau)$ is the scalar charge whose physical meaning will be made clear later.
The sub-indices in $t$ and $\tau$ specify the respective events $x,\;y$ and
$z$. That $t_{x}$ must be equal to $t_{z}$ on the LHS  of eq. (\ref{rof}) is a
consequence of the deltas on its RHS and of the constraint (\ref{hcone}).
Initially, it is assumed that both ${\dot q}\equiv\frac{dq}{d\tau}$ and ${\ddot
q}\equiv\frac{d{\dot q}}{d\tau}$ exist and that they may be
non null.
The field eq. (\ref{dfe}) is solved by
\be
\l{phif}
\phi_{f}(x,\tau)=\int d^5yG_{f}(x-y,\tau_{x}-\tau_{y})\rho(y,\tau_{y})
\ee
with
\be
\l{boxGf}
\eta^{\mu\nu}\nabla_{\mu}\nabla_{\nu}G(x,\tau)=\delta^{(5)}(x).
\ee
The discrete Green's function associated to the Klein-Gordon operator is
given\cite{paperI} by
\be
\label{pr9}
G_{f}(x,\tau)=\frac{1}{2}\theta(bf^4t)\theta(b\tau)\delta(\tau+
f.x),\;\;\;\;{\vec x}_{{\hbox{\tiny T}}}=0,
\ee
where $b =\pm1,$ and $\theta (x)$ is the Heaviside function, $\theta(x\ge0)=1$
and $\theta(x<0)=0.$ The labels {\tiny L} and {\tiny T} are used as an
indication of, respectively,  longitudinal and transversal with respect to the
space part of $f$: ${\vec f}.{\vec x}_{\hbox{\tiny T}}=0$ and $ x_{{\hbox{\tiny
L}}}=\frac{{\vec f}.{\vec x}}{|{\vec f}|}$.

Remarkably $G_{f}(x,\tau)$ does not depend on anything outside its support, the
fibre $f$, as stressed by the append ${\vec x}_{{\hbox{\tiny T}}}=0$. One could
retroactively use this knowledge in the action (\ref{da1}) for rewriting it as
\be
\l{sf3}
S_{f}=\int d^{5}x\delta^{(2)}({\vec x}_{\hbox{\tiny
T}}){\Big\{}{1\over2}\eta^{\mu\nu}\nabla_{\mu}\phi_{f}\nabla_{\nu}\phi_{f}-\chi\phi_{f}(x,\tau)\rho(x,\tau){\Big\}},
\ee
just for underlining that the fibre $f$ induces a conformally invariant (1+1)
theory of massive and massless fields, embedded in a (3+1) theory, as
generically discussed in paper I. Actually, the factor $\delta^{(2)}({\vec
x}_{\hbox{\tiny T}})$ is an output of the actions (\ref{da1}) or (\ref{ar}) (it
is not necessary to put it in there by hand) and it can never be incorporated
as a factor in the definition (\ref{pr9}) of $G_{f}(x,\tau)$, except under
integration sign as in eqs.
(\ref{phif}) and (\ref{sf3}).

Then one could, just formally, use
\be
\l{dsf}
\rho_{[f]}(x-z,\tau_{x}-\tau_{z})=q(\tau)\delta(\tau_{x}-\tau_{z})\delta(t_{x}-t_{z})\delta(x_{\hbox{\tiny L}}-z_{\hbox{\tiny L}}),
\ee
where $\rho_{[f]}$ represents 
the source density $\rho$ stripped of its explicit ${\vec x}_{\hbox {\tiny
T}}$-dependence, for reducing the action to
\be
\l{21}
S_{f}=\int d\tau_{x}dt_{x}dx_{\hbox{\tiny
L}}{\Big\{}{1\over2}\eta^{\mu\nu}\nabla_{\mu}\phi_{f}\nabla_{\nu}\phi_{f}-\chi\phi_{f}(x,\tau)\rho_{[f]}(x,\tau){\Big\}},
\ee
by just omitting the irrelevant transversal coordinates. Eq. (\ref{da1}) then,
after its output eq. (\ref{pr9}), is formally equivalent to eq. (\ref{21}).

The solutions from eq. (\ref{usfe}) for a point source are well known massless
spherical waves propagating (forwards and backwards in time) on a lightcone in
contradistinction to the solutions (\ref{pr9}) that are, massive or massless,
point signals propagating (always forwards in time) on a straight line that
happens to be a generator of the hypercone (\ref{hcone}), the one that is
tangent to $f$. Being massive or massless is determined by $\tau$ being
constant or not, as discussed in paper I. For a massive field, its mass and its
time-like four velocity are hidden behind a lightlike $f$ and a non-constant
$\tau$; they only become explicit after the passage from discrete to continuous
fields. But as it will be made clear in Section V, there is no point on
considering a massive discrete scalar field because any discrete scalar field
must be associated to the gravitational field of general relativity. So massive
discrete scalar fields will not be considered here any further.

Using the eqs. (\ref{pr9}) and (\ref{rof}) with $b=+1$, $f^4\ge1$ that is, for
the emitted field, one has
$$
\phi_{f}(x,\tau_{x})=\chi\int d^{5}y
\theta(t_{x}-t_{y})\theta(\tau_{x}-\tau_{y})\delta[\tau_{x}-\tau_{y}+
f.(x-y)]q(\tau_{z})\delta^{4}(y-z)=$$
\be
=\chi\int d\tau_{y}
\theta(t_{x}-t_{y})\theta(\tau_{x}-\tau_{y})\delta[\tau_{x}-\tau_{y}+
f.(x-y)]q(\tau_{z}),
\ee
where an extra factor 2 accounts for a change of normalization with respect to
eq. (\ref{pr9}) as  the annihilated field (the integration over the future
lightcone) is being excluded. Then,
\be
\l{Af0}
\phi_{f}(x_{{\hbox{\tiny L}}},{\vec x}_{{\hbox{\tiny T}}}={\vec
z}_{{\hbox{\tiny T}}},t_{x},\tau_{x}=\tau_{z})=\chi
\theta(t_{x}-t_{z})\theta(\tau_{x}-\tau_{z})q(\tau_{z}){\Big |}_{f.(x-z)=0}
\ee
or for short, just
\be
\l{Af1}
\phi_{f}(x,\tau)=\chi q(\tau)\theta(t)\theta(\tau){\Big |}_{f}.
\ee

$\nabla\theta(t)$ and $\nabla\theta(\tau)$ do not contribute \cite{paperI} to
$\nabla \phi_{f},$ except at $x=z(\tau),$ as a further consequence of the field
constraints. So, for $t>0$ and (therefore) $\tau\geq0$  one can write just
\be
\l{AfV}
\phi_{f}(x,\tau_{x})=q(\tau_{z}){\Big |}_{f}
\ee
\be
\l{dAf}
\nabla_{\nu}\phi_{f}=-f_{\nu}{\dot q}{\Big |}_{f}
\ee
With eq. (\ref{dAf}) in eq. (\ref{det}) one has
\be
\l{tsf1}
T^{\mu\nu}_{f}(x,\tau_{x})=f^{\mu}f^{\nu}{\dot q}^{2}{\Big |}_{f}
\ee
The field four-momentum, given by $\int T^{\mu\nu}n_{\nu}d\sigma$ for a
continuous field, is reduced, thanks to the omission of the transversal
coordinates, to
\be
\l{psf}
p^{\mu}_{f}=T^{\mu\nu}_{f}n_{\nu}=f^{\mu}{\dot q}^{2}{\Big |}_{f}
\ee
where $n$ is a space like four vector \cite{paperI} such that $n.f=1$. The
conservation of the energy-momentum content of $\phi_{f}$ is assured by
$f^2=0,$
\be
\l{emc}
\nabla_{\mu}T^{\mu\nu}_{f}=-2f_{\mu}f^{\mu}f^{\nu}{\dot q}{\ddot q}{\Big
|}_{f}=0.
\ee
It is justified naming $\phi_{f}$ a discrete field because although being a
field it is not null at just one space point at a time; but it is not a
distribution, a Dirac delta function, as it is everywhere and always finite.
Its differentiability, in the sense of having space and time derivatives, is
however assured by its dependence on $\tau$, a known, supposedly continuous
spacetime function. It is indeed a new concept of field, a very peculiar one,
discrete and differentiable; it is just a finite point-like spacetime
deformation projected on a null direction, with a well defined and everywhere
conserved energy-momentum. It is this discreteness in a field that allows the
union of wave-like and particle-like properties in a same physical object,
which implies \cite{hep-th/9911233} finiteness and no spurious degree of
freedom (uniqueness of solutions).

 \section{Discrete physics}

{}From eq. (\ref{AfV}) we see that the field $\phi_{f}$ is given, essentially,
by the charge at its retarded time, i.e. the amount of scalar charge at $z$,
the event of its creation.  $\phi_{f}$ has a physical meaning in the sense of
having a content of energy and momentum when and only when ${\dot q}\not=0$.
So, the emission or the absorption of a scalar field is, respectively,
consequence or cause of a change in the amount of scalar charge on its source.
This is so because emitting or absorbing a scalar field requires a change in
the state of its source which is so poor of structure that has nothing else to
change but itself, and this is fundamental for determining the scalar charge
nature. The picture becomes clearer after recalling that  one  is dealing with
discrete field and discrete interactions implying that the change in the state
of a field source occurs at an isolated event.  $q(\tau)$ is not a continuous
function:
\be
\l{dV}
q(\tau):=\sum_{i}q_{\tau_{i+1}}{\bar\theta}(\tau_{i+1}-\tau){\bar\theta}(\tau-\tau_{i}),
\ee
where
\be
\l{dth}
{\bar\theta}(x)=\cases{1, &if $x>0$;\cr 1/2,& if $x=0$;\cr 0,& if $x<0$,\cr}
\ee
and the sum is over the interaction points on the source world  line,
$i=1,2,3\dots$.
\be
q(\tau)=\cases{q_{\tau_{j}}, & if $\tau_{j}<\tau<\tau_{j+1}$;\cr
		&\cr
		\frac{q_{\tau_{j-1}}+q_{\tau_{j}}}{2}, & if $\tau=\tau_{j}$;\cr
		&\cr
		q_{\tau_{j-1}}, & if $\tau_{j-1}<\tau<\tau_{j}$,\cr}
\ee
as indicated in  the graph of the Figure 1.
\vglue13cm
\hglue-2.0cm
\begin{minipage}[]{7.0cm}
\parbox[t]{5.0cm}{
\begin{figure}
\vglue-14cm
\epsfxsize=300pt
\epsfbox{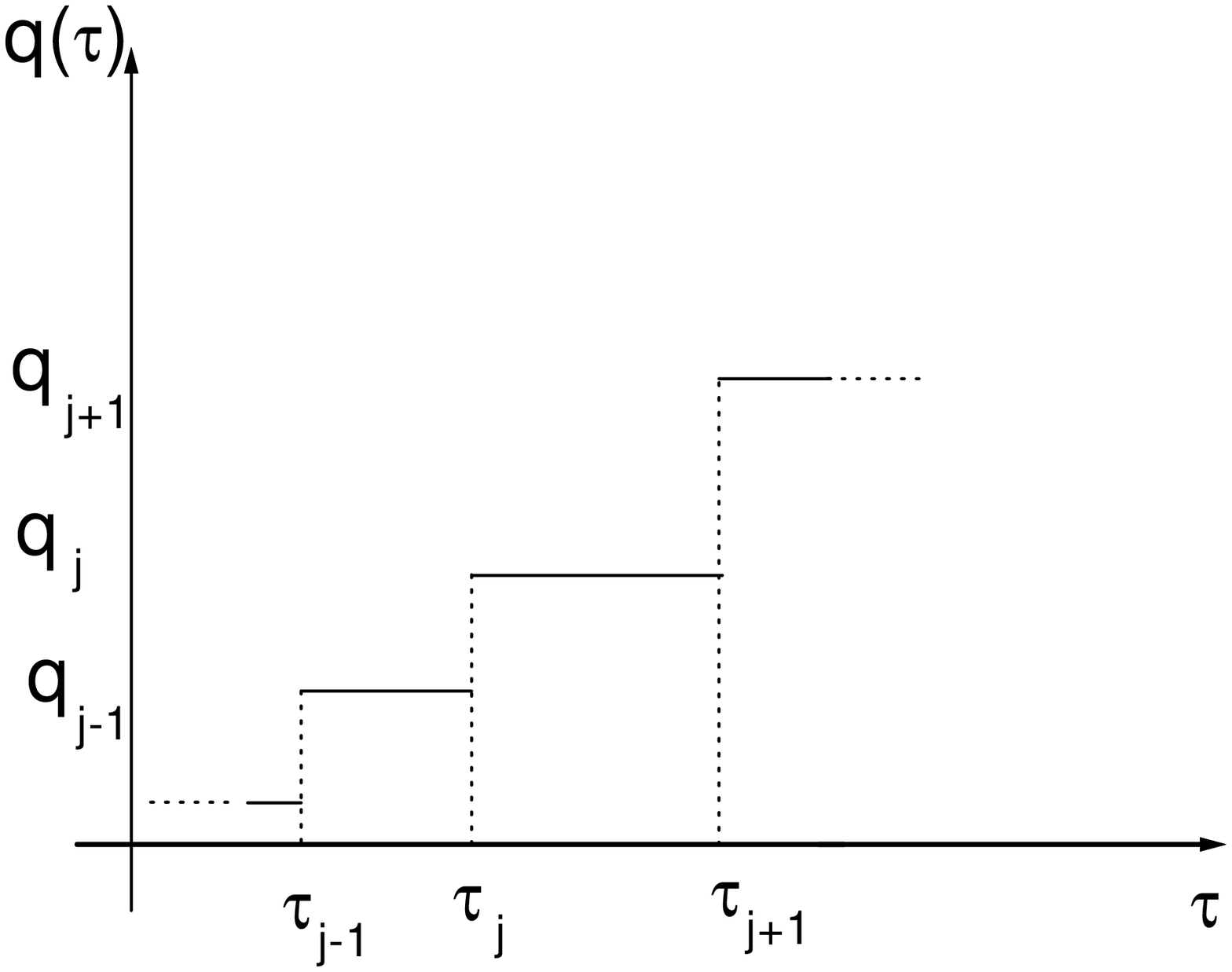}
\end{figure}}
\end{minipage}\hfill
\vglue-7cm
\hglue-1.0cm
\begin{minipage}[]{7.0cm}
\begin{figure}
\hglue9.0cm
\l{f3a}
\parbox[t]{5.0cm}{\vglue-6cm
\caption{Discrete changes on a discrete scalar charge along its world  line. A
discrete scalar charge is so poor of structure that there is nothing else to
change but itself. There is change in the state of a scalar source only at the
interaction points on its world  line which is labelled by its proper time. If
only the (discrete) interaction points are relevant the proper time may be
treated as a discrete variable. In the limit of a world  line densely packed of
interaction points a continuous graph is a good approximation.}}
\end{figure}
\end{minipage}

\vglue-1cm

The change in the state of the scalar source is not null only at the
interaction points and so, rigourously, it cannot  be defined as a time
derivative, as there is no continuous variation, just a sudden finite change.
The naive use of
\be
\l{dasdd}
{\dot q}=q(\tau)\delta(\tau-\tau_{z}),
\ee
would be just an insistence on an unappropriate continuous formalism, besides
artificially introducing infinities where there is none. It means that one must
replace time derivatives by finite differences
\be
\l{finitedif}
{\dot q}(\tau)\Rightarrow \cases{\Delta q_{\tau_{j}} & if $\tau=\tau_{j}$;\cr
				0 & if $\tau\neq\tau_{j}$,\cr}
\ee
and a proper-time integration by a sum over the interaction points on the
charge world  line.
The existence and meaning of any physical property that corresponds to a time
derivative must be reconsidered at this fundamental level. Velocity ($v)$
exists as a piecewise smoothly continuous function (discontinuous at the
interaction points). Acceleration ($a$) and derivative concepts like force
($F$), power ($P$), etc rigorously do not exist. We must deal with finite
differences, respectively, the  sudden changes of velocity ($v$), momentum
($p$) and energy ($E$):
\be
\l{reciped}
\cases{a\Rightarrow\Delta v\cr
F\Longrightarrow\Delta p \cr
P\Rightarrow\Delta E\cr}
\ee
So, calculus (integration and differentiation) in a discrete-interaction
context becomes useless for a rigourous description of fundamental physical
processes. But in practice such a strictly discrete calculus is not always
feasible. What effectively counts is the scale determined by $\Delta\tau_{j}$,
the time interval between two consecutive interaction events, face the accuracy
of the measuring apparatus. The question is if $\Delta\tau_{j}$ is large enough
to be detectable, or how accurate is the measuring apparatus used to detect it.
The density of interaction points on the world  line of a given point charge is
proportional to the number of point charges with which it interacts. Let one
consider the most favourable case of a system made of just two point charges.
As the argument is supposedly valid for all fundamental interactions one can
take the hydrogen atom in its ground state for consideration, treating the
proton as if it were also a fundamental point particle. The order of scale of
$\Delta\tau_{j}$ for an electron in the ground state of a hydrogen atom is
given then by the Bohr radius  divided by the speed of light
$$\Delta\tau_{j}\sim10^{-18}s$$
which corresponds to a number of ${\pi\over\alpha}\sim400$ interactions per
period ($\alpha$ is the fine-structure constant) or $\sim10^{10}$
interactions/cm. So, the electron world  line is so densely packed with
interaction events that one can, in an effectively good descriptions for most
of the cases, replace the graph of the Figure 1 by a continuously smooth curve.
The validity of calculus in physics is then fully re-established as a
consequence of the limitations of the measuring apparatus. The Wigner's
questions\cite{Wigner} about the unexpected  effectiveness of mathematics in
the physical description of the world is recalled.  The answer lies on the huge
number of point sources in interaction (a sufficient condition), the large
value of the speed of light and the small (in a manly scale) size of atomic and
sub-atomic systems, which indirectly is a consequence of h, the Planck
constant.


Although $\Delta\tau_{j}$ may not be measurable, at least with the present
technology, the discrete formalism is justified not for replacing the
continuous one where it is best, which       is confirmed by high precision
experiments \cite{Nakanish?} but mostly for defining and understanding its
limitations. There are, besides this very generic justification, many instances
of one-interaction-event phenomena, like the Compton effect, particle decay,
radiation emission from bound-state systems, etc, where discrete interactions
are the natural and the more appropriate approach. These are, of course, all
examples of quantum phenomena, but primarily because quantum here implies
discreteness. A discrete field with discrete interactions not only requires a
proper quantum context  but it makes easier and natural the passage from the
classical one. This is being discussed in the companion paper\cite{paperIII}.

\begin{center}
Discrete-continuous transition
\end{center}

 It would be interesting to have a framework where this change from continuous
to discrete interaction and vice-versa could be formally realized in a simple
and direct way.
One can deal with them considering the behaviour under a derivative operator of
 ${\bar\theta}(\tau)$ which is the mathematical description of the interaction
discreteness. Then one must require that, symbolically
\be
\l{dk}
\frac{\partial}{\partial\tau}{\bar{\theta}}(\tau-\tau_{i}):=\delta_{\tau\tau_{i}},
\ee
with $\delta_{\tau\tau_{i}}$ the Kronecker delta
\be
\delta_{\tau\tau_{i}}=\cases{1, &if $\tau=\tau_{i}$;\cr
	&\cr
	0, & if $\tau\ne\tau_{i}$,\cr}
\ee
with the meaning that at the points where the LHS of eq. (\ref{dk}) is not
null, which are the only relevant ones,  $\tau$ must be treated as a discrete
variable and that the operator $\frac{\partial}{\partial\tau}$ must be seen as
(or replaced by) just a sudden increment $\Delta$ and not as the limit of the
quotient of two increments.

Then with such a convention one has from eq. (\ref{dV}) that
\be
\l{nabla}
\nabla_{\nu}q(\tau){\Big
|}_{f}=-f_{\nu}\sum_{i}q_{\tau_{i}}\{{\bar\theta}(\tau_{i+1}-\tau)\delta_{\tau\tau_{i}}-\delta_{\tau_{i+1}\tau}{\bar\theta}(\tau-\tau_{i})\}:=-f_{\nu}{\dot q}(\tau),
\ee
which implies that ${\dot q}(\tau)$ is null when $z(\tau)$ is not a point of
interaction on the charge world  line. For such an interaction point $\tau_{j}$
one has
\be
{\dot q}(\tau_{j})
=q_{\tau_{j}}{\bar\theta}(\tau_{j+1}-\tau_{j})-q_{\tau_{j-1}}{\bar\theta}(\tau_{j}-\tau_{j-1})=q_{\tau_{j}}-q_{\tau_{j-1}}
\ee
or, generically
\be
\l{ac}
{\dot q}(\tau)=\cases{\Delta q_{i}=q_{\tau_{i}}-q_{\tau_{i-1}} & for
$\tau=\tau_{i}$;\cr
		&\cr
		0 & for $\tau\ne\tau_{i}$,\cr}
\ee
and, from the middle term of eq. (\ref{nabla})
\be
\nabla^{f}_{\sigma}\nabla^{f}_{\nu}q(\tau)=-2f_{\sigma}f_{\nu}\sum_{i}q_{\tau_{i}}\delta_{\tau\tau_{1}}\delta_{\tau_{i+1}\tau}=0.
\ee
 In eq. (\ref{nabla}) one has summation over the vertices and only these points
on the world  line contribute. That is why one has to define  eq. (\ref{dk}).
In a limit where this summation may be approximated by an integration the
Kronecker delta may be replaced by a Dirac delta function and then one may have
eq. (\ref{dasdd}) as a good operational approximation to eq. (\ref{ac}).\\
Then
\be
\l{dAj}
\phi_{f}(x)=q(\tau){\Big |}_{f}=\cases{q_{\tau_{j+1}}{\Big |}_{f} &if
$\tau_{j}<\tau_{ret}<\tau_{j+1}$\cr
		&\cr
		&\cr
		 \frac{q_{\tau_{j+1}}+q_{\tau_{j}}}{2}{\Big |}_{f}& if
$\tau_{ret}=\tau_{j}$\cr}
\ee
and
\be
\l{dAj}
\nabla_{\mu}\phi_{f}(x)=-f_{\mu}\Delta q(\tau){\Big
|}_{f}=\cases{-f_{\mu}(q_{\tau_{j+1}}-q_{\tau_{j}}){\Big |}_{f} &if
$\tau_{ret}=\tau_{j}$\cr
		&\cr
		 0& if $\tau_{ret}\not=\tau_{j}$\cr}
\ee
The field $\phi_{f}(x,\tau)$ is just like an instantaneous picture of its
source at its retarded time. A travelling picture.
If $z(\tau_{ret})$ is not a point of change in the source state, $\phi_{f}(x)$
does not describe a real field; its energy tensor is null. A real field always
corresponds to a sudden change in its source state at its retarded time. If
there is no change the field is not real, in the sense of having zero energy
and zero momentum. Having no physical attribute it corresponds to the pure
``gauge field" of the continuous formalism.

\section{Scalar field and general relativity}

It takes an external agent on the scalar source to cause a change $\Delta q$ on
its charge $q$; a positive $\Delta q$ means that a scalar field
$\phi_{f}(x,\tau)$ has been absorbed whereas a negative one means an emission.
Therefore, a discrete scalar field carries itself a charge $\Delta q$ and can,
consequently, interact with other charge carriers and be a source or a sink for
other discrete scalar fields. It carries a bit of its very source, a scalar
charge; it is an abelian charged field.
 On the other hand a new look at equations (\ref{psf}) and (\ref{ac}) reveals
that $(\Delta q_{j})^2$ describes the energy content of the field. So, the
source of a discrete scalar field is any physical object endowed with energy
which corresponds then to the scalar charge.  Energy, of course, is a component
of a four-vector and not a Lorentz scalar. Its four-vector character comes from
the $f^{\mu}$ factor in eq. (\ref{psf}): the energy of $\phi_{f}(x,\tau)$ is
the fourth component of the current of its squared scalar charge. The scalar
charge conservation is therefore assured  by and reduced to the conservation of
energy and momentum given by eq. (\ref{emc}).  Considering the relativistic
mass-energy relation this implies that the discrete scalar field satisfies the
Principle of Equivalence and that all physical objects interact with the scalar
field through its energy-tensor.  This is a form of the Principle of
Universality of gravitational interaction, introduced by
Moshinski\cite{Moshinsky}. So, $\phi_{f}(x,\tau)$must necessarily be connected
to the gravitational field. Having necessarily energy for source implies on an
important consequence of uniqueness, of excluding the existence of any other
distinct fundamental discrete scalar field as it must necessarily be taken as
the gravitational field\footnote{There would be no point on assuming that a
same charge could be the source of two or more distinct fields, having besides
the same characteristics}. Moreover, as energy is not a scalar, the symmetry
between discrete fields and  sources, both taken as fundamental fields, implies
also that there is no fundamental scalar source representing an elementary
field; it must be a scalar function of a non-scalar fundamental field, like the
trace of an energy tensor, for example. This lets then explicit a known
symmetry of nature: the four fundamental interactions are described by gauge
fields having  for sources vector currents (schematically, $j=qv$), of their
respective charges $q$, including gravity since the energy tensor is just a
current of its charge, the four-vector momentum. So, this symmetry is not
broken with gravity being a second-rank tensor field.\\
This possible physical interpretation is compatible with the General Theory of
Relativity, according to the work done in references
\cite{gr-qc/9801040,gr-qc/9903071}, where a discrete gravitational field
defined by
\be
\l{gmn}
g_{\mu\nu}^{f}(x)=\eta_{\mu\nu}-f_{\mu}f_{\nu}\chi\phi_{f}(x,\tau),
\ee
as a point deformation in a Minkowski spacetime, propagating on a null
direction $f$, upon an integration on $f$, in the sense of eq. (\ref{dAf}),
reproduces the standard continuous solutions. That gravity be either totally
\cite{inMTW} or partially \cite{Fierz,Darmour} described by a scalar
(continuous) field is an old idea\cite{FierzePauli,Jordan,Brans-Dicke}, but eq.
(\ref{gmn}) implies on regarding gravity as being ultimately described by the
scalar field in a metric theory. With the metric in this form the Einstein's
field equations
\be
\l{Eeq}
R_{\mu\nu}-{1\over2}g_{\mu\nu}R=\chi T_{\mu\nu}
\ee
is reduced \cite{gr-qc/9801040} to
\be
\l{rEeq}
f_{\mu}f_{\nu}\eta^{\alpha\beta}\nabla_{\alpha}\nabla_{\beta}\phi_{f}(x,\tau)=\chi T_{\mu\nu},
\ee
as the gauge condition used in \cite{gr-qc/9801040}
\be
f^{\mu}\nabla_{\mu}\phi_{f}(x,\tau)=0
\ee
becomes an identity after eq. (\ref{dAf}), as $f^2=0$.

Inherent to discrete fields, irrespective of their tensor or spinor character,
is the  implicit  conservation of their sources as a consequence of their
(discrete fields) very definition, as discussed in Section V of paper I.
Whereas $T^{\mu\nu};_{\mu}=0$  is assured by the symmetry of the Einstein
tensor on the LHS of eq. (\ref{Eeq}), in eq. (\ref{rEeq}) it is just a
consequence (see eq.(\ref{emc})) of eq. (\ref{dAf}). This symmetry of the
Einstein tensor is in this way similar to the one of the Maxwell tensor that
assures charge conservation in the standard continuous-field formalism but that
is a consequence of extended causality (discrete-field definition) and Lorentz
symmetry \cite{paperIII,hep-th/9911233} in a discrete-field approach.

The eq. (\ref{gmn}) recalls an old derivation \cite{Feynman} of the field
equations of general relativity by consistent re-iteration of
\be
\l{reit}
g_{\mu\nu}(x)=\eta_{\mu\nu}+\chi h(x)_{\mu\nu},
\ee
as solution from a gauge invariant wave equation for the field $g_{\mu\nu}(x)$
in a Minkowski spacetime. The non-linearity of the Einstein's equations comes
from contribution of all higher orders in $h_{\mu\nu}$ to $g_{\mu\nu}(x)$.
Therefore, the results obtained in reference \cite{gr-qc/9801040} imply that if
$h_{\mu\nu}$ is ultimately a discrete scalar field on a fibre $f$
$$h_{\mu\nu}=f_{\mu}f_{\nu}\phi_{f}(x,\tau)$$ there is no higher order
contribution essentially because $f^2=0$. A discrete field has no
self-interaction, a consequence of its definition (\ref{df}) and that is
explicitly exhibited in its Green's function (\ref{pr9}). Discrete fields are
solutions from linear  equations. Whereas this is true for $g^{f}_{\mu\nu}$ of
eq. (\ref{gmn}) it is not for its $f$-average $g_{\mu\nu}$ of eq. (\ref{reit}).
The non-linearity of general relativity appears here then as a consequence of
the averaging process of eq. (\ref{s}) that effectively smears the discrete
field over the lightcone \cite{gr-qc/9801040,gr-qc/9903071}.

On the other hand  the energy tensor in eq. (\ref{rEeq}) must be traceless,
also a consequence of $f^2=0$. This recalls an old known problem in standard
field theory that comes when a massless theory is taken as the
$(m\rightarrow0)-$limit of a massive-field theory
\cite{BoulwareeDeser,Deser,Van-Dam,Feldman,Fierz}, but for a discrete field, in
contradistinction, a traceless tensor does not necessarily mean a massless
source \cite{paperI}.
The wave equation (\ref{rEeq}) must be preceded by some careful qualifications,
however. A discrete field requires a discrete source. The source in eq.
(\ref{rEeq}) must be treated as a discrete set of point sources
$T^{f}_{\mu\nu}(x,\tau)$ for which $f^{\mu}T^{f}_{\mu\nu}(x,\tau)=0$. This
implies that there is no exterior solution for a discrete gravitational field,
only vacuum solutions. Any interior continuous solution must be seen then as an
approximation for a densely packed set of point sources. From the discrete
vacuum solution of eq. (\ref{rEeq}) one can, in principle, with an integration
over its $f$-parameters, obtain any continuous vacuum solution of
an imposed chosen symmetry\footnote{From the superposition of the discrete
fields of a spherical distribution of massless dust one retrieves the Vaydia
metric\cite{unpublished}.}\cite{gr-qc/9801040}. This justifies, up to a certain
point, not regarding the RHS of eq. (\ref{gmn}) as just the first two terms of
a series of possible contributions from higher rank tensors.
 Even for a massive point-source, however, being itself a discrete field,
$T^{f}_{\mu\nu}$ cannot be expressed in terms of its mass and of its actual
four-velocity $v$. A traceless $T^f_{\mu\nu}(x,\tau)$ with
$f^{\mu}T^{f}_{\mu\nu}(x,\tau)=0$ does not necessarily represent a massless
source nor $f$ represents its four-velocity, as discussed in Section V of paper
I.

The geometrical description of gravity as the curvature of a pseudo-Riemannian
spacetime has its validity always assured as an absolutely good approximation
due to the high density of interaction points in any real measurement, as
discussed in the previous section.
\section{Conclusions}
The thesis that fundamental interactions are discrete has being developed. If
this is the case there is no really compelling reason for excluding gravity
from such a unifying idea. It is necessary to emphasize, however, that no one
is proposing the replacement of general relativity by a discrete scalar field
theory of gravity. The point is that for gravity it is extremely doubtful that
it could make an experimentally detectable difference. It is not a novelty
that, considering the small strength of the gravitational coupling this
interaction is irrelevant for physical systems involving relatively few
fundamental elements. Observe that even a gravitational Aharanov-Bohm-like
experiment \cite{Sakurainaoda!} would require the gravitational field of a
macroscopically large object, like the Earth. The sufficient condition for a
high density of interaction points is assured and justifies continuous
descriptions of gravity, of which general relativity is undoubtedly the best
proposal \cite{Darmour}. Moreover  the undectability of discrete gravity is
tantamount to the unobservability of the Minkowski spacetime. At this level  of
approximation the Minkowski spacetime becomes the local tangent space of a
curved space-time and $f$ is a generator of the local hypercone in the tangent
space. This would lead to full general relativity in accordance to a general
uniqueness result \cite{MTW,Visser} that any metric theory with field equations
linear in second derivatives of the metric, without higher-order derivatives in
the field equations, satisfying the Newtonian limit for weak fields and without
any prior geometry must be exactly Einstein gravity itself. This reminds  again
the already mentioned \cite{Feynman} derivation of general relativity from flat
spacetime but now with the distinctive aspect that the effective Riemann space
comes not from a consistency requirement but as an approximation validated by a
recognized limitation on the present experimental capacity. On the other hand,
the knowledge of a supposedly true discrete character of all fundamental
interactions, gravity included, is a permanent reminder of the limits of such a
continuous approximative description, irrespective of how good their results
fit with experiments. The idea of an essential continuity of any physical
interaction allows unlimited speculations that will always go beyond any level
of possible experimental verifications which brings then the risk of not being
able of distinguishing  the reign of possibly experimentally-grounded
scientific research  from plain philosophical speculation or even just fiction.
A discrete gravitational interaction, even if not experimentally detectable,
still for a long time to come, may just make sense of existing theories as it
has historically happened with all new discreteness introduced in the past,
like the ideas of molecules, atomic transitions, and quarks, for example.


\begin{thebibliography}{10}
\bibitem {paperI}M.M. de Souza, {\it Conformally symmetric massive discrete
fields}, submitted to publication. hep-th/0006237.
\bibitem {hep-th/9610028} M. M. de Souza, J. of Phys. A: Math. Gen. 30
(1997)6565-6585. hep-th/9610028
\bibitem{Wigner} E. P. Wigner, {\it The Unreasonable Effectiveness of
Mathematics in Natural Sciences} reprinted in {\it The Collected Works of
Eugene Paul Wigner}, Vol VI, part B, J. Mehra ed., Springer Verlag,
Berlin(1995); R.Jackiw {\it The Unreasonable Effectiveness of Quantum Field
Theory} hep-th/9602122.
\bibitem{paperIII} In preparation.
\bibitem {gr-qc/9801040} M. M. de Souza, Robson N. Silveira,  Class. \& and
Quantum Gravity, vol 16, 619(1999).
\bibitem{Polchinski}J. Polchinski, {\it String theory. Vol. I}, Cambridge Univ.
Press, Cambrige,(1998).
\bibitem{Jackson} D. Jackson {\it Classical Electrodynamics},2nd ed., chaps. 14
and 17,
John Wiley {\&} Sons, New York, NY(1975).
\bibitem{inMTW}G. Nordstrom, Ann. Phys. (Germ.)   42, 533(1913); see Section
17.6 of reference \cite{MTW}.
\bibitem{MTW}C.W. Misner, K.S. Thorne, J.H. Wheeler, {\it Gravitation},
Freeman,S. Francisco(1973).
\bibitem{FierzePauli}M. Fierz, W. Pauli, Proc. Roy. Soc., 173,211(1939).
\bibitem{Jordan}P.Jordan, Nature 164, 637(1949).
\bibitem{Brans-Dicke}C. Brans, R.H. Dicke, Phys. Rev. 124, 925(1961),{\it
ibid.}125,2194(1961).
\bibitem{hep-th/9911233} M.M. de Souza,{\it {Discrete gauge fields}}
hep-th/9911233
\bibitem{BoulwareeDeser}D.G. Boulware and S. Deser, Phys. Rev. D6,3368(1972).
\bibitem{Deser}S. Deser, J. Gen. Rel. Grav.,1,9(1970).
\bibitem{Van-Dam}H. Van Dam, M. Veltman, Nucl. Phys.,B22,397(1970).
\bibitem{Feldman}G. Feldman,{\it Classical Electromagnetic and Gravitational
Field Theories as Limits of Massive Quantum Theories} in {\it The Uncertainty
Principle and Foundations of Quantum Mechanics; a fifty years survey},W.C.
Price and S.S. Chissick (eds.)- John Wiley \& Sons, London(1977).
\bibitem{Fierz}M. Fierz, Helv. Acta 29, 128(1956).
\bibitem{Darmour}C.M.Will, {\it Theory and Experiment in Gravitational
Physics}, 2nd. ed. Cambridge University Press, Cambridge, 1993; T.Darmour, {\it
Experimental tests of Relativistic Gravity}, gr-qc/9904057.
\bibitem{Sakurainaoda!}R. Colella, A.W. Overhauser, S.A Werner. Phys. Rev.
Lett. 34, 14722(1975). See also the references in G.Z. Adunas, E.
Rodriguez-Milla, D.V. Ahluwalia, {\it Probing quantum violations of the
equivalent principle} gr-qc/0060022.
\bibitem{unpublished}M. M. de Souza, unpublished.
\bibitem{Nakanish?} T. Kinoshita and D.R. Yennie in{\it Quantum
Electrodynamics} T. Kinoshita (ed.), World Scientific, Singapore(1990).
\bibitem{Feynman}W. Wyss,Helv. Phys. Acta,38(65)469;S. Deser,
Gen.Rel.Grav,1(1970) 9.
\bibitem{Moshinsky}Moshinsky M., Phys. Rev. v. 80,514(1950); R.P.Feynman,
F.B.Morinigo and W.G. Wagner,{\it Feynman Lectures on
Gravitation},B.Hatfield(ed),Addison-Wesley,Reading(1995).
\bibitem{gr-qc/9903071}M. M. de Souza, Robson N. Silveira, {\it Gauge vs
Coulomb and the cosmological mass-deficit problem}, gr-qc/9903071.
\bibitem{Visser}M. Visser {\it Mass for the graviton}, gr-qc/9705051
\end{thebibliography}
\end{document}